# Nanostructured thin films of indium oxide nanocrystals confined in alumina matrixes

A. Bouifoulen [a,b], M. Edely [a], N. Errien [a], A. Kassiba [a,*], A. Outzourhit [b], M. Makowska-Janusik [c], N. Gautier [d], L. Lajaunie [d], A. Oueriagli [b]

[a] *Laboratoire de Physique de l'Etat Condensé, UMR-CNRS 6087, Institut de recherche IRIM2F-FR-CNRS 2575-Université du Maine, Avenue Olivier Messiaen, 72085 Le Mans Cedex 9, France*
[b] *Laboratoire de Physique du Solide et Couches Minces, Faculté des Sciences Semlalia, Université Cadi Ayyad, B. P. 2390, Marrakech 40000, Maroc*
[c] *Institute of Physics, Jan Dlugosz University, 13/15 Al.Armii Krajowej, 42-200 Czestochowa, Poland*
[d] *Institut des Matériaux Jean Rouxel (IMN), Université de Nantes-CNRS, UMR 6502, 2 rue de la Houssinière, BP 32229, 44322 Nantes Cedex, France*

## abstract

Nanocrystals of indium oxide ($In_2O_3$) with sizes below 10 nm were prepared in alumina matrixes by using a co-pulverization method. The used substrates such as borosilicate glasses or (100) silicon as well as the substrate temperatures during the deposition process were modified and their effects characterized on the structural and physical properties of alumina-$In_2O_3$ films. Complementary investigation methods including X-ray diffraction, optical transmittance in the range 250-1100 nm and transmission electron microscopy were used to analyze the nanostructured films. The crystalline order, morphology and optical responses were monitored as function of the deposition parameters and the post-synthesis annealing. The optimal conditions were found and allow realizing suitable nanostructured films with a major crystalline order of cubic phase for the $In_2O_3$ nanocrystals. The optical properties of the films were analyzed and the key parameters such as direct and indirect band gaps were evaluated as function of the synthesis conditions and the crystalline quality of the films.
.

## 1. Introduction

The need of innovating integrated optoelectronics has motivated the interest for nanostructured materials which allow stable and efficient light emission in selective wavelength windows. Thus, a wide variety of quantum dots based architectures were realized and their optical features were characterized for potential applications. As examples, cadmium telluride nanodots embedded in three-dimensional photonic crystals exhibit modified emission spectrum [1] while, a self assembly of quantum dots (QD) exhibits original architectures with electronic and optical features depending on the organisation of QD [2]. Other nanostructures based on Ge nanocrystals embedded in a silica matrix or silicon carbide nanoparticles give rise to photoluminescence responses [3,4].

In this context, indium oxide ($In_2O_3$) is an attractive wide band-gap semiconductor with promising properties for optoelectronics [5,6], solar cells [4,7] or for photocatalysts [8]. Beyond synthesis routes developed to obtain thin films with or without doping, $In_2O_3$ nanostructures such as nanotubes [9,10], nanobelts [11-13], nanofibers [14,15], wires [16-23], and nanoparticles [24-26] have been widely emphasized to extend their technological applications.

In this work, an original synthesis approach was developed to obtain guest-host nanostructures where $In_2O_3$ nanocrystals are distributed and confined in alumina matrix. The choice of such structures is motivated by the possibility to explore the photoluminescence of $In_2O_3$ nanocrystals without any optical limitations from the host alumina matrix. Indeed, alumina ($Al_2O_3$) is one of the most suitable materials for optical coatings in a wide spectral region [27]. Therefore, it is considered to be the most promising candidates able to be associated to $In_2O_3$ in order to fine-tune the optical responses of the overall guest-host films.

The main challenge of this work deals with the synthesis of suitable nanostructured thin films based on In2O3 nanocrystals regularly embedded in alumina by using co-evaporation of reactants in rf-sputtering deposition chamber. The synthesis parameters are related to the substrate natures (borosilicate, quartz, crystalline silicon) and the deposition conditions such as the sputtering power, argon partial pressure as well as the temperature of the substrates. Additionally, the structural features were characterized as a function of the annealing cycles on the composite films performed under oxidized atmospheres. The different parameters were judiciously varied in order to improve the organization of the nanostructured films. In this respect, complementary characterization methods were conjugated to shed light on the structure, morphology and optical peculiarities of the films as function of deposition conditions and annealing treatments. Thus, transmission electron microscopy (TEM), X-ray diffraction and reflectivity, as well as UV-Vis absorption experiments were carried out and a systematic evaluation of the characteristics of the films was monitored as function of the experimental conditions. Choosing the optimal parameters has contributed to improve the crystallization of $In_2O_3$ nanoparticles in the cubic phase. This achievement opens the possibility to explore the targeted applications related to photoluminescence responses.

---

* Corresponding author. Tel.: +33 2 43 83 35 12.
  *E-mail address:* kassiba@univ-lemans.fr (A. Kassiba).

2. Experimental details

The reactant precursors such as alumina and $In_2O_3$ nanopowders used for the deposition were purchased from Aldrich. Appropriate targets were realized and the $In_2O_3$-Alumina nanostructured films were prepared by rf-sputtering method using co-deposition processes from a target composed by one large alumina pellet (33 mm) and several small pellets (5 mm) of indium oxide. The later were hot-pressed from mechanically made pellets followed by an iso-static pressure foaming and a sintering at temperatures up to 1300 °C. This treatment is required to make hardly maintained $In_2O_3$ pellets, appropriate for the rf-sputtering process. Silicon wafers and borosilicate glasses (BK7) were mainly used as substrates depending on the required structural or optical characterizations for the synthesized films. The deposition conditions include different substrate temperatures up to 500 °C while an almost constant sputtering rate of deposition about 4.5 nm/min is achieved by using a rf-sputtering power fixed at about 50 W and an argon pressure of $10^{-2}$ mbar. Furthermore, by using a deposition time about 60 min, the obtained film thickness was in the range 270-320 nm evaluated by optical interferometer measurements. After the deposition process, annealing was also carried out in a separate chamber at temperatures up to 400 °C under air during 3 hours and the effect was characterized from the crystalline quality of the nano-objects.

Structural, optical and morphology characteristics of the nanostructured films were investigated by complementary techniques. Thus, the film thickness was measured by using an optical interferometer (FilmeTrics Model 205-0070). The crystalline phases involved in the synthesized films were identified by X-ray diffraction (XRD) using a Philips X-pert diffractometer with a Cu-Kα radiation and employing a scan rate of 0.02°/s in the scattering angular range (2θ) 10-70°. The optical transmittance was measured by UV-visible spectrophotometer (DH-2000-BAL) in the wavelength of 250-1100 nm leading also to absorption coefficients as a function of the synthesis conditions. Cross section morphologies of nanostructured films deposited on silicon were investigated by using a high resolution transmission electron microscopy (HRTEM). In this aim, samples were prepared in the cross-section geometry (X-TEM). They were mechanically thinned using a tripod polisher down to 10 μm, and then ion milled in a GATAN PIPS apparatus at low energy (2.5 keV Ar) and low incidence (± 8 ) to minimize irradiation damage. The prepared samples were studied by High Resolution Transmission Electron Microscopy (HRTEM) using a Hitachi H9000NAR microscope (300 kV, $LaB_6$, point to point resolution= 0.18 nm).

3. Results and discussion

*3.1. Structural features probed by X-ray diffraction*

Nanostructured alumina-$In_2O_3$ films were deposited on (100) silicon and BK7 substrates maintained at different temperatures (20 °C, 250 °C, 400 °C, and 500 °C). X-ray diffraction (XRD) analyses carried out on the initial reactant $In_2O_3$ powder (Fig. 1(A)) revealed a pure cubic phase and will serve as a reference structure for the deposited films. XRD lines of the as-deposited and annealed films are shown in Fig. 1(B) and (C) as a function of the substrate nature. However, while glass substrates are required notably for optical measurements, their main drawback for XRD investigations lies in the superimposed background signal to the XRD lines of the samples. In some extent, this contribution causes an ambiguous assignment of the XRD results. As an example, the XRD patterns (C) show a broad hump-like feature around 2θ=28° which can be either attributed to the amorphous nature of the BK7 substrate, to the small nanocrystallites or to an amorphous organization of the films [28] as well. Comparative analysis by using different substrates and different treatments are then required for a precise insight on the crystalline structures involved in Alumine-$In_2O_3$ films. Thus, analyses of XRD patterns suggest that the microstructural evolution and transformation kinetics of the $In_2O_3$ from amorphous to crystalline phase are highly dependent on the used substrate temperatures and on the annealing temperatures. It is worth noting that in all cases, the alumina matrix did not show any crystalline order and it seems amorphous-like whatever the adopted synthesis conditions. In the as-deposited films at room temperature and at 250 °C, $In_2O_3$ structures are amorphous whereas the 400 °C deposited film exhibits quasi-crystalline structure. For the film deposited on silicon, the XRD patterns exhibit mainly well resolved peaks identified as (2 2 2), (4 0 0) and (4 4 0) reflections of a crystalline $In_2O_3$ cubic structure (I a −3 space group with lattice parameter a=b=c= 10,117 A°). However, it is well known that preferential orientations of crystalline growth can occur on the synthesized films and the phenomena can be enhanced if columnar or filamentary arrangements are involved. These microstructures are drastically dependant on the deposition conditions of the films and on the nature of the substrates [29-31]. In such cases, the intensity of the XRD patterns can vary widely from that of the bulk material. The reported XRD studies confirm the existence of preferential growth orientation along the (2 2 2) direction when the substrate temperature (400 °C) is used to improve the crystalline order of the $In_2O_3$ structures. In fact, in the case of the nanostructured film deposited on BK7, the low intensity of the (6 6 2) and (4 4 0) XRD lines compared to the (4 0 0) one, characterizes the role of the substrate on the stabilized crystalline structure. Moreover, the increase of the substrate temperature seems to enhance the kinetic energy and hence the mobility of the species on the substrate surface leading to an improved crystalline order of the deposited films. The performed XRD experiments point out the crucial role of substrate temperature and particularly the one at 400 °C which represents the critical temperature required for the improvement of the crystalline order of the synthesized films.

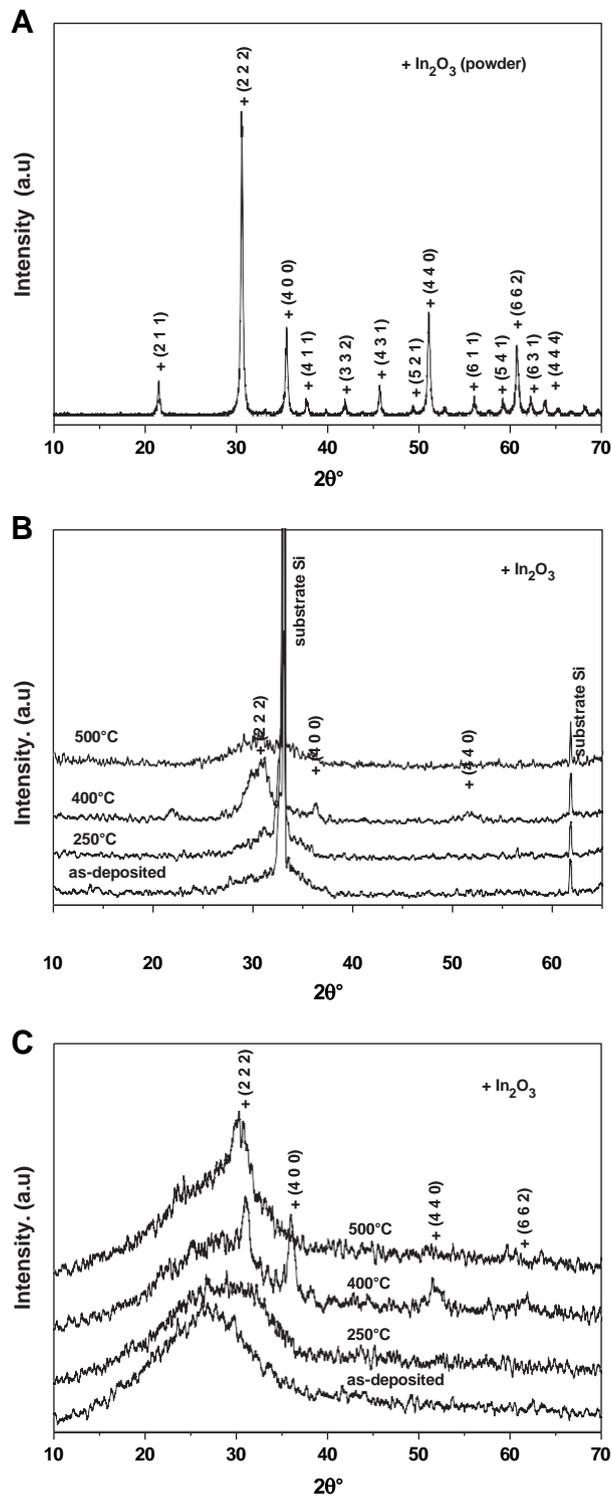

Fig. 1. (A) X-ray diffraction (XRD) pattern of $In_2O_3$ nanopowders with the diameter of the particles in the range 10-100 nm. (B) XRD patterns of the as-synthesized alumina-$In_2O_3$ on a silicon substrate formed at different substrate temperatures with the assignment of the XRD lines with respect to the cubic structure of $In_2O_3$. (C) XRD patterns of the as-synthesized alumina-$In_2O_3$ films deposited on borosilicate (BK7) substrates maintained during the synthesis at different temperatures reported on the plots.

### 3.2. Morphology of the $In_2O_3$ clusters in the host alumina

The organization of the nanostructured films was examined by TEM investigations. With regard to the difference in the electronic density between indium and aluminum, the expected good contrast is thought to discriminate between the host alumina matrix and the $In_2O_3$ clusters. Fig. 2(A) shows a typical X-TEM micrograph of the synthesized films at 400 °C. Onto the Si substrate, a thin film of 270 nm thick is observed, in good agreement with other determinations of film thickness by optical means. The microstructure of the film presents a columnar contrast due to the organization of $In_2O_3$ nanoparticles in columnar forms resulting mainly from the synthesis method by rf-sputtering. This morphology has been already observed for example in Cu-based nanocomposite silica films [32]. The amorphous alumina matrix contains elongated free spaces between columns of alumina. The growth of $In_2O_3$ nanoparticles takes place in the free volumes involved in the alumina film with a typical size in the range of 2-10 nm. The coalescence of $In_2O_3$ clusters in this Alumina free space gives rise to more or less elongated objects perpendicular to the surface with an effective diameter of crystalline regions about 40 nm. Fig. 2(B) displays X-HRTEM micrograph of the sample synthesized at 400 °C and confirms the crystalline nature of the nanoparticles in good agreement with XRD observations. The inset of Fig. 2(B) shows the Fast-Fourrier Transform (FFT) image performed on the area highlighted by a black square in Fig. 2(B). A detailed analysis of the inter-reticular distances derived from this FFT and a comparison with simulated diffraction patterns given by JEMS software [33] have shown that this investigated $In_2O_3$ nanocrystal presents the cubic crystallographic structure seen along the [111] direction. Similar analyses have been performed on several nanocrystals. Even if for some clusters it was not obvious to discriminate between rhomboedral and cubic phases, it should be noted that only the cubic structure has been clearly evidenced by HRTEM.

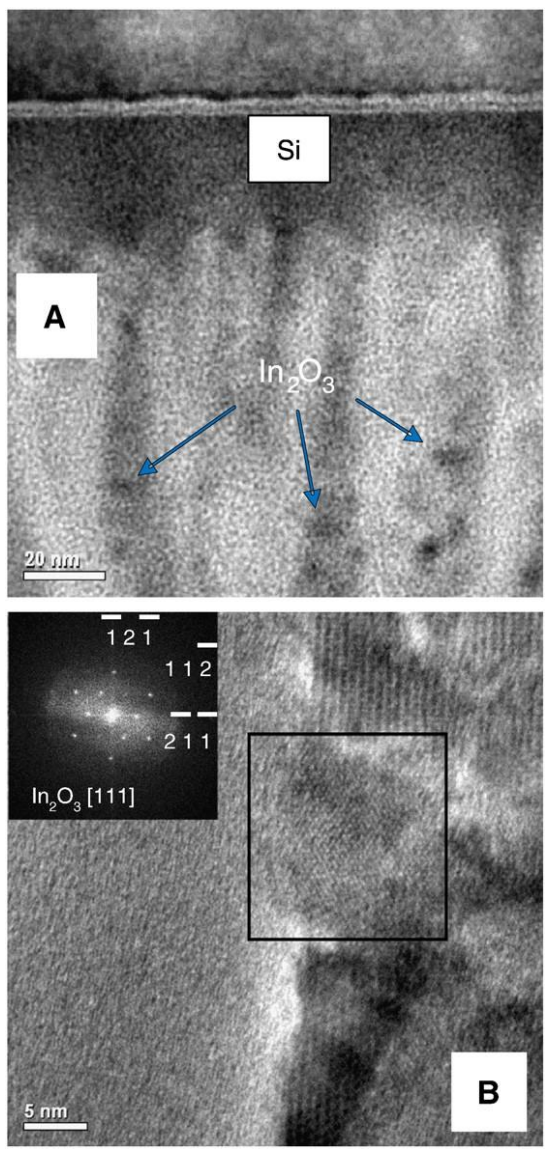

Fig. 2. TEM images of a thin film synthesized at 400 °C with isolated $In_2O_3$ nanoparticles aligned in columnar blocks (A). HRTEM micrograph (B) obtained on one cluster of nanoparticles. The inset in panel (B) shows the FFT performed on the area highlighted by a black square. The indexation of the FFT reveals that the investigated $In_2O_3$ nanoparticle crystallized in the cubic phase seen along the [111] direction.

### 3.3. Optical properties of nanostructured films

#### 3.3.1. General features

Comparative UV-Vis absorption curves are summarized in Fig. 3 for several thin films synthesized by using different substrate temperatures. Transmittance and absorption coefficients are shown respectively in Fig. 3(A) and (B) for different substrate temperatures. The ripples in the spectrum result from the interference fringes due to suitable film thickness giving rise to interferences features [34]. The pronounced changes on the optical properties correlate with the incorporation of $In_2O_3$ in alumina matrix and with the crystallization degree being improved by changing the substrate temperature. The absorption coefficients deduced from experimental curves, indicate a clear shift of the adsorption edges to shorter wavelength with increasing the substrate temperature up to 400 °C [35]. However, a higher temperature (500 °C) seems to induce a reverse evolution caused by a low crystalline order demonstrated by XRD characterizations.

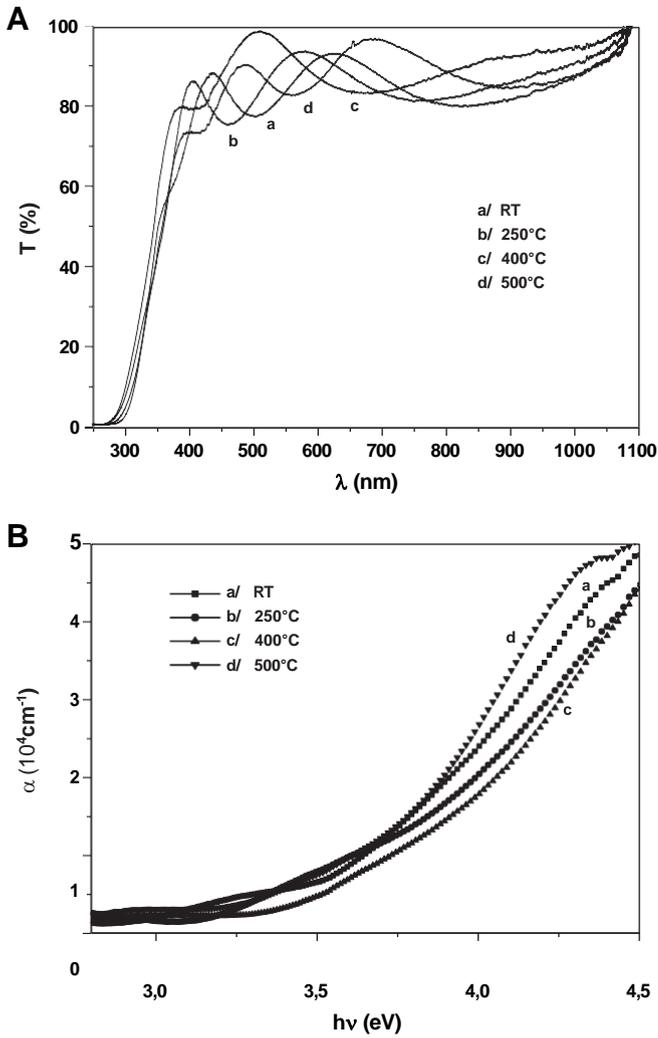

Fig. 3. (A) Optical transmission versus the wavelength of light for different substrate temperatures given on the graph. (B) Typical variation of absorption coefficient with photon energy for alumina-In$_2$O$_3$ films with different BK7 substrate temperatures.

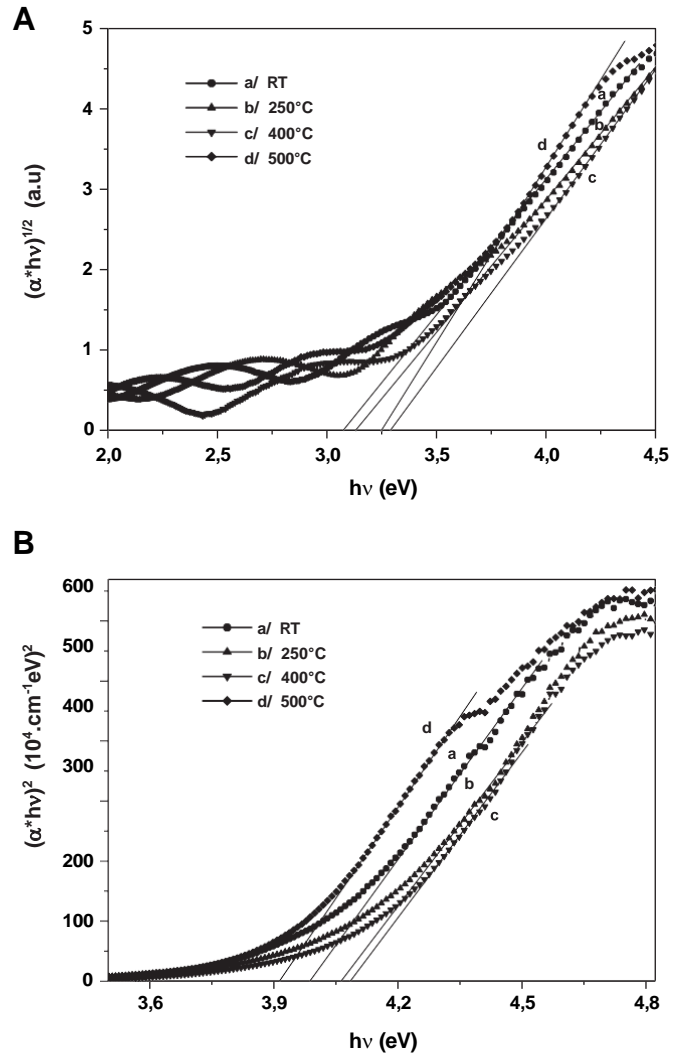

Fig. 4. (A) Representation of the curve $(\alpha h\nu)^{1/2}$ versus $h\nu$ and (B) graph of $(\alpha h\nu)^2$ versus $h\nu$ for alumina-In$_2$O$_3$ deposited at different substrate temperatures. An extrapolation of the linear regions of the plots determines the band-gap characteristics of the In$_2$O$_3$ structures as discussed in the text.

### 3.3.2. Estimation of alumina-In$_2$O$_3$ nanostructure band gaps

The absorption coefficient versus photon energy can be used to estimate the band gap of the deposited films. The possibility that direct and indirect bad gaps exist can be pointed out by using appropriate representations of the experimental UV-Vis absorption. Additionally, insights on the origin of optical transitions may be obtained from the approach hereafter outlined. The direct and indirect band gaps of the alumina-In$_2$O$_3$ films may roughly be estimated by plotting the $(\alpha h\nu)^n$ versus the photon energy ($h\nu$). Depending on the exponent "$n$," several data can be obtained. Indeed, $n = 2$ is required to evaluate the allowed direct band gap while $n = 1/2$ is used for the estimation of an indirect one. Other situations such as $n = 1/3$ or $n = 2/3$ are respectively relevant for forbidden indirect and forbidden direct optical transitions. All these band gaps are estimated by suitable extrapolation of the linear region of the plot toward low energies [36]. Fig. 4(A) and (B) exemplify these plots for the alumina-In2O3 films synthesized by using different substrate temperatures [37,38].

Extrapolating the linear parts of the $(\alpha h\nu)^{1/2}$ versus $h\nu$ plots (Fig. 4(A), $n = 1/2$) gives an indirect band gap about 3.08, 3.13, 3.29 and 3.25 eV. The same procedure for the $(\alpha h\nu)^2$ plots (Fig. 4(B), $n = 2$) gives a direct band gap about 3.98, 4.05, 4.08 and 3.91 eV for the films deposited respectively at RT, 250 °C, 400 °C and 500 °C. Thus, the indirect band gap estimated in the present work for the nanostructured In$_2$O$_3$ incorporated in alumina matrix is intermediate between the same characteristics determined for pure alumina and pure In$_2$O$_3$ [39-41] (Fig. 5, Table 1). These results point out the possibility to modulate the effective band gap of the nanostructured films by using different ratios of indium oxide clusters in the host alumina matrixes.

### 4. Conclusions
5.

Guest-host nanostructured films based on alumina as host matrix and nanocrystals of In$_2$O$_3$ were synthesized with the aim to fine tune the optical band gap by adjusting the nanocrystal size and the crystalline order. XRD and TEM experiments showed that the improvement of the crystalline degree of alumina-In$_2$O$_3$ nanostructures requires the use of different substrates maintained at temperatures between room temperature and 500 °C. The critical deposition temperature about 400 °C improves the crystalline quality of In2O3 nanoclusters with a mainly two involved polytypes such as cubic and rhombohedral. TEM observations point out the occurrence of columnar morphology of the films with free volumes in the range 2-10 nm. The growth of In$_2$O$_3$ nanocrystals is marked by coalescence to a more or less elongated objects perpendicular to the surface with an effective diameter of crystalline regions about 40 nm. The optical properties of the films were investigated and the key parameters such as direct and indirect band gaps were evaluated. Thus, the allowed-indirect and allowed-direct optical band gaps were found to vary respectively from 3.08 to 3.29 eV and from 3.98 to 4.08 eV, depending on the temperature of the substrates and then on the crystalline quality of the films. The realization of suitable nanostructures based on $In_2O_3$ nanocrystals confined in host amorphous alumina matrix is well achieved from the carried out work. Investigations of the photoluminescence features of such nanostructures supported by numerical simulations are now under scope.

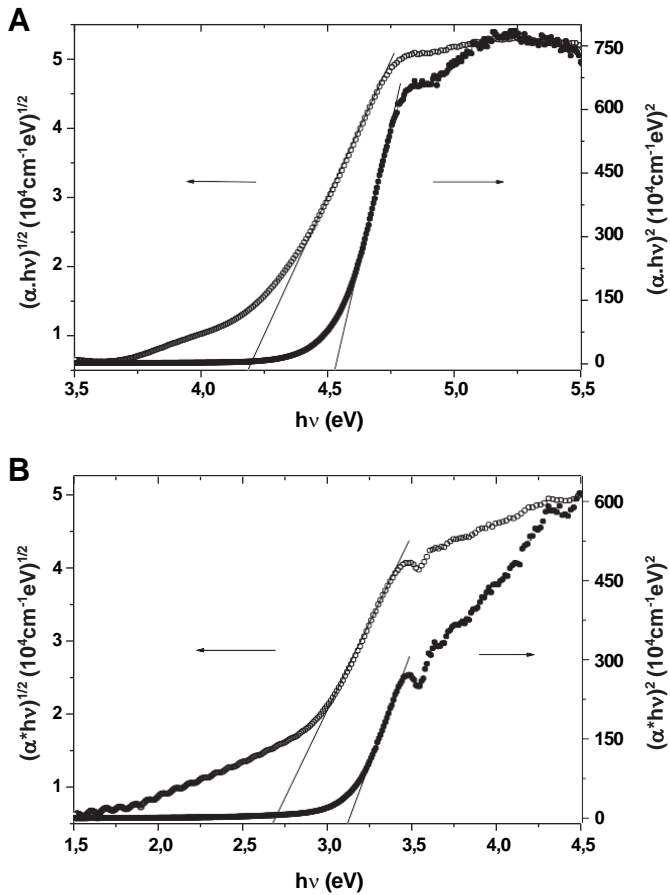

Fig. 5. Graphs of (A) $(\alpha h\nu)^{1/2}$ versus $h\nu$ and (B) $(\alpha h\nu)^2$ versus $h\nu$ for bare alumina matrix and bare $In_2O_3$ thin films. An extrapolation of the linear regions of the plots determines the corresponding direct and indirect band gaps.

Table 1
Direct and indirect band gap in alumina and $In_2O_3$ thin films synthesized in the same conditions as the nanostructured films based on nanocrystals of $In_2O_3$ incorporated in alumina matrix.

|  | Indirect gap (eV) | Direct gap (eV) |
|---|---|---|
| (A) $Al_2O_3$ | 4.18 | 4.52 |
| (B) $In_2O_3$ [39–41] | 2.67 | 3.12 |

Acknowledgements

A part of the work was realized under the France-Morocco cooperation program VOLUBILIS with the financial support from the "Comité inter-universitaire Franco-Marocain."